\documentclass[]{aastex3}

\usepackage{amssymb}  
\usepackage{amsthm}
\usepackage{amsmath}
\usepackage{hyperref}
\usepackage{cleveref}

\shorttitle{The properties of high redshift galaxies}
\shortauthors{L.A. Garc\'ia}

\begin{document}

\title{The properties of high redshift galaxies}

\author[0000-0003-1235-794X]{Luz \'Angela Garc\'ia}
\affiliation{Universidad ECCI. Cra. 19 No. 49-20, Bogot\'a, Colombia, C\'odigo Postal 111311}
\email{lgarciap@ecci.edu.co}

\begin{abstract}
This work studies the connection between the first galaxies and their hosting dark matter halos in the early Universe when Reionization is concluding. Our numerical models (already presented in an earlier study) trace the star formation history at $z =$ 4 - 8, the galaxy stellar mass function, the stellar-to-halo mass distribution, and other high redshift galaxies statistics. All these predictions are consistent with observations to date and other high-resolution cosmological simulations. A key finding of this work is the robust estimate for the cosmic star formation history (through the implementation of galaxy and supernova winds and atomic and molecular cooling processes) and self-consistent chemical pollution of the intergalactic medium. The theoretical models are compatible with a faint-end slope of the galaxy luminosity function of $\alpha =$ -2 at the end of the Reionization.
\end{abstract}


\section{Introduction}
The formation and evolution of galaxies at high redshift strongly determined the progression of the Epoch of Reionization (EoR). The hierarchical model for the structure formation provides a scenario where small dark matter halos ($\sim$ 10$^{6}M_{\odot}$) at $z \sim$ 30 reached a critical temperature to agglomerate baryons, and formed stars \citep{tegmark1997}, initially metal-free (POPIII). However, these first stars evolved and consumed their fuel quickly and produced the first supernova explosions. With them, the chemical pollution of the intergalactic medium (IGM) began \citep{matteucci2005}. Consequently, a new generation of stars (POPII) formed the first black holes in the Universe. Interestingly, metals in POP II stars supplied an additional cooling source; therefore, these structures were less massive and had longer lives than metal-free stars. The interaction of the collapsed systems in the Universe and the IGM is described through feedback models that account for winds that spread out chemical elements and vary the ionization state of free Hydrogen.\newline

Using Hubble Space Telescope imaging, \citet{robertson2015} measure the abundance and luminosity distribution of early galaxies and provide a constraint for the observed star formation rate (SFR). However, the largest compilation to date for the SFR is presented by \citet{madau2014}. Later works by \citet{oesch2015,mcleod2015,finkelstein2015,song2016,oesch2018,ishigaki2018,rachana2019}; among others, have complemented our survey of galaxies at early times, pushing the detections to very faint objects at higher redshifts \citep[$z \sim$ 11;][]{robertson2021}.\newline 

On the other hand, theoretical models have a three-fold purpose: i) match the observations available for the synthetic galaxies and their environment; ii) provide a physical description of the star formation process at all times, and iii) interpolate -when possible- the progression of galaxies and their properties in time. There are multiple models that reproduce the star formation history of galaxies at early times, focusing on specific physical processes, among them: \textsc{Angus} \citep[\textit{AustraliaN} \textsc{Gadget-3} \textit{early Universe Simulations},][]{tescari2014}; the \textsc{Renaissance} suite \citep[with the AMR code \textsc{Enzo},][]{oshea2015}; \textsc{Eagle}
\citep[Evolution and Assembly of GaLaxies and their Environments,][]{crain2015,schaye2015}; CROC \citep[Cosmic Reionization on Computers,][]{zhu2020}; the \textsc{Obelisk} simulation \citep{trebitsch2020}; L-Galaxies 2020 \citep{henriques2020,yates2021a}; \textsc{Flares} \citep[First Light And Reionisation Epoch Simulations,][]{lovell2021,vijayan2021}; \textsc{Astraeus} 
\citep[semi-numerical rAdiative tranSfer coupling of galaxy formaTion and Reionization in N-body dArk mattEr simUlationS,][]{hutter2020}.\newline

With the advent of the James Webb Space Telescope (JWST) shortly, we will reach an unprecedented understanding of the first light and the Reionization, as well as the assembly of galaxies \citep{gardner2006}. The expectation is that JWST will observe galaxies out to $z >$ 12 and even at $ z \sim $ 15 (depending on their brightness). It will provide a uniform census of galaxies in the redshift range of 7 - 12 and extend the observed cosmic star formation evolution in a level not achievable by Hubble \citep{finkelstein2016}. In addition, JWST will: detect stars with very low metallicity (10$^{-3}Z_{\odot}$), set constraints of the top-heavy IMF, provide an estimate on the escape fraction of ionizing photons of galaxies, and allow for a robust investigation of the UV luminosity function at high redshift (\citealt{atek2015} found a steep faint-end slope at $z \sim$ 7). Theoretical works as \textsc{UniverseMachine} by \citet{behroozi2020}, IllustrisTNG with dust modeling \citep{vogelsberger2020} or \citet{williams2018} -that creates a mock catalog of galaxy populations from the UV to the near-infrared- anticipate future observations from the instrument and provide a realistic (yet conservative) forecast of the galaxy assembly at redshift up to 15.\newline

The paper is presented as follows: Section~\ref{sec:2} gives an overview of our numerical simulations and the physical modules implemented to recreate the structure formation at high redshift. Also, we briefly present a description of other numerical simulations considered to evaluate the performance of our models. Section~\ref{sec:3} shows a series of statistics for galaxy and halo properties in the simulations at 4 $<z<$ 8. We also compare our theoretical predictions with observational data available to-date and the largest compilation of high-resolution cosmological models in this redshift range. We establish a connection between the galaxy properties and their hosting dark matter halo and show a self-consistent chemical enrichment in the models. Section~\ref{sec:4} discusses future scenarios where our conclusions can be tested, as well as the strengths and caveats of our models. Finally, Section~\ref{sec:5} summarizes the findings and conclusions of this study. 

\section{The numerical simulations}
\label{sec:2}

This work relies on a set of high-resolution hydrodynamical simulations at high redshift (4 $ < z < $ 8), initially presented in \citet{garcia2017b}. The model is based on a customized version of the smoothed particle hydrodynamics (SPH) code {\textsc{Gadget-3}} \citep{springel2005c}, with a spatially flat cosmology $\Lambda$CDM model and cosmological pa\-ra\-me\-ters from \citet{planck2015}, $\Omega_{0m}=$ 0.307, $\Omega_{0b}=$ 0.049, $\Omega_{\Lambda}=$ 0.693, $n_s=$ 0.967, $H_0=$ 67.74 km s$^{-1}$Mpc$^{-1}$ (or $h =$ 0.6774) and $\sigma_8 =$ 0.816. A summary of the numerical simulations is shown in Table~\ref{table_sims_2}.\newline

The numerical models are complemented with an algorithm that identifies collapsed structures, so-called parallel Friends-of-Friends (FoF), and a parallel {\small{SUBFIND}} algorithm to identify substructures within FoF halos.\newline

The mechanism for which star-forming gas particles turn into star-type particles by a stochastic process was first proposed by \citet{katz1996} and later, discussed in \citet{springel2003,tornatore2007}. \newline
Thus, our model produces self-consistently chemical enrichment based on the stochastic scheme for star formation. The module follows the evolution of Hydrogen, Helium, and nine elements up to iron, delivered from SNIa and SNII and intermediate-mass stars. Importantly, stars with masses $m \leq$ 40 $M_{\sun}$ explode as supernovae before turning into a black hole. Conversely, stars above such threshold collapse into a black hole without experiencing the supernova stage, contributing to the feedback process, but not to the chemical evolution in the simulations. The overall state of a simple stellar population depends on the lifetime function \citep{padovani1993}, the stellar yields, and the initial mass function (IMF). \newline

\begin{table}
\centering
\caption{\footnotesize{Overview of the simulations used in the paper. The first column corresponds to the name of the run, the second one, the box size. The third column is the comoving gravitational softening length. Columns 4 and 5: gas and DM particle masses. Note that all runs have the same initial number of gas and DM particles ($2\times 512^3$). The acronyms MDW and EDW stand for momentum- and energy-driven winds feedback prescriptions, and \textbf{mol} at the end of the run's name indicates the presence of low-temperature metal and molecular cooling. The fiducial models is highlighted in bold in the first row: Ch 18 512 MDW (highlighted in bold).}}
\label{table_sims_2}
\vspace{0.5cm}
\resizebox{0.98\textwidth}{!}{%
\begin{tabular}{|lcccc|} 
  \hline
  Simulation &  Box size & Comoving softening & $M_{\text{gas}}$ & $M_{\text{DM}}$ \\
 &  (cMpc/$h$) & (ckpc/$h$) & ($\times$ 10$^{5}M_{\sun}$/$h$) & ($\times$ 10$^{6}M_{\sun}$/$h$) \\ \hline
\textbf{Ch 18 512 MDW}  & \textbf{18} & \textbf{1.5} & \textbf{5.86} & \textbf{3.12} \\ 
Ch 18 512 MDW mol & 18 & 1.5 & 5.86 & 3.12 \\ 
Ch 18 512 EDW  & 18 & 1.5 & 5.86 & 3.12 \\ 
Ch 18 512 EDW mol & 18 & 1.5 & 5.86 & 3.12 \\ 
Ch 12 512 MDW mol & 12 & 1.0 & 1.74 & 0.925 \\ 
Ch 25 512 MDW mol & 25 & 2.0 & 15.73 & 8.48 \\ 
\hline
\end{tabular}
}
\end{table}

The stellar yields account for the amount of metals released by each source during the stellar evolution: SNIa \citep{thielemann2003}, SNII \citep{woosleyweaver1995} and low- and intermediate-mass stars. Moreover, the solar metallicity layers follow results from \citet{asplund2009}.\newline

The cooling processes that allow the gas to form stars include atomic, metal-line cooling \citep{wiersma2009}, as well as low-temperature cooling by molecules and metals \citep{maio2007,maio2015}.\newline

On the other hand, this work builds on a multi-sloped IMF \citep{chabrier2003} that accounts for massive POP II and, to some extent, to POP III stars, which significantly contribute to the first stages of the star formation processes and the Hydrogen Reionization.\newline

Our numerical simulations implement galactic winds to regulate the star formation process, the dispersion of metals from galaxies to the intergalactic medium (IGM), and prevent overcooling of the gas \citep{springel2003}. Such feedback mechanisms expel material and balance the temperature among neighbor gas particles, allowing physical processes to occur. There are two kinetic supernova--driven winds considered in this work: energy- \citep[EDW; ][]{springel2003} and momentum-driven winds \citep[MDW; ][]{puchwein2013}, and AGN feedback \citep{springel2005b,fabjan2010,planelles2013}. The latter type of feedback is essential at low redshift ($z \sim$ 2) when massive halos are more numerous and massive.\newline

In this work, both EDW and MDW supernova outflows are implemented, with a fixed fiducial velocity $v_{\text{fid}} =$ 600 km/s. The main assumption in the former prescription for the winds is the proportionality between the star formation rate $\dot{M}_{\star}$ and the mass-loss rate due to winds $\dot{M}_w$, through the relation $\dot{M}_w=\eta \dot{M}_{\star}$. The factor $\eta$ is defined as the wind mass loading factor and quantifies the efficiency of the wind to expel material out of the source cell.\newline

The kinetic energy of the wind and the halo circular velocity allow us to establish a numerical relation between wind mass--loading factor $\eta$ and $v_w$ \footnote{ See \citet{garcia2017b} for the complete derivation of this expression.}: 
\begin{equation}\label{eta2}
\eta=2 \times \left(\frac{v_{\text{fid}}}{v_w}\right)^2.
\end{equation}

Nonetheless, \citet{puchwein2013} show that the star formation rate $\dot{M}_{\star}$ and the mass expelled by supernova winds $\dot{M}_w$ do not necessarily have a linear relation. Instead, a more natural assumption would be a mathematical relationship for the star formation rate of the galaxy and the winds' momentum flux. In such case, $\eta \propto v_w^{-1}$:
\begin{equation}\label{eta22}
\eta=2 \times \frac{v_{\text{fid}}}{v_w}.
\end{equation}
\noindent It is worth mentioning that the wind velocity $v_w$ has the same functional form as in the energy-driven winds feedback. Yet, their efficiencies $\eta$ behave in distinctive ways because of the scaling with $v_w$.\newline

On the other hand, different authors have shown that AGN feedback is critical to regulating the star formation rate history, gas accretion, stellar evolution, and metal enrichment when the Universe has evolved for 10 billion years (i.e., $z \sim$ 2), at the peak of the star formation and consequently, the most significant quasar activity in the history of the Universe. \newline 
Nevertheless, \citet{tescari2014} present an extensive discussion of the negligible effect of AGN feedback at the redshifts of interest of this work. Two main factors determine that AGN do not play an essential role at high redshift: i) the galaxies are still experiencing their first stages of star formation; hence, very few super-massive black holes have formed at this time; ii) dark matter halos are still growing by mergers; thus, AGNs (if existing) are rare, and so, their feedback mechanisms. 

\subsection*{Other numerical simulations}
In order to convey a successful comparison of the predictions from this set of simulations with current theoretical models, we briefly summarize the main features of each of the mock cosmological boxes, highlighting their resolution and box sizes.

\begin{itemize}
\item \textsc{UniverseMachine}\newline
As described in \citet{behroozi2020}, the \textsc{UniverseMachine} is based on the Very Small MultiDark-Planck (VSMDPL) simulation, a modified version of \textsc{GADGET-
2} with a flat $\Lambda$CDM model and $h =$ 0.68. This model was run from $z =$ 150 to 0, in a cube size of 160 cMpc/$h$ with 3840$^{3}$ particles, allowing it to reach a gas mass of 9.1 $\times$ 10$^{6}$M$_{\odot}$, a numerical resolution of 2 ckpc/$h$ at $z > 1$ and resolved dark matter halos with 100 particles (i.e., above 10$^{9}$M$_{\odot}$). The latter property of the \textsc{UniverseMachine} makes it suitable to appropriately describe halos at high redshift and foresee the characteristics of undetected galaxies at the Epoch of Reionization. 
\item L-Galaxies 2020\newline
This semianalytical model of galaxy evolution \citep{henriques2020} is run on the \textsc{Millennium-II} simulations in a box of $\sim$ 96.1 Mpc/$h$ side. It only contains dark matter particles (the baryonic physics is implemented through effective modules that are easily adapted), reaching a broad coverage at the cosmological level. The dark matter particles have a mass resolution of 7.7$\times$ 10$^{6}$M$_{\odot}$/$h$. The model is scaled to the Planck 2013 cosmology with $h =$ 0.673. L-Galaxies 2020 currently has two distributions: 
\textbf{Default model} or \textbf{DM}, first described in \citet{henriques2020}, assumes that 70\% of the metal content is released by supernova is instantly mixed with the local interstellar medium (ISM) before being expelled out of galaxies via SN winds. Instead, the \textbf{Modified model -MM-} \citep{yates2021a} adopts a chemical pollution prescription where up to 90\% of metals produced in supernova explosions are moved directly to the circumgalactic medium (CGM) without passing by the ISM. These two complementary scenarios cover a wide range of CGM enrichment schemes, likely to occur in real galaxies.
\item \textsc{Eagle}\newline
Evolution and Assembly of GaLaxies and their Environments (\textsc{Eagle}) is a  hydrodynamical suite of cosmological simulations run in a modified version of \textsc{Gadget-
3} \citep{crain2015,schaye2015}. The model's main strength is the galaxy growth and evolution, and it includes similar prescriptions and modules as the ones implemented in our numerical simulations. For this paper, we will only focus on their largest volume 'L100N1504' cube, with 67 (Mpc/$h$)$^{3}$ box-size and 2$\times$ 1504$^{3}$ particles (dark matter $+$ gas). The initial mass is 1.2 (6.6) $\times$ 10$^{6}$M$_{\odot}$/$h$ for baryons (and dark matter). The assumed cosmology is Planck 2013 with $h =$ 0.6777, and their supernova feedback is EDW. 
\item TNG100\newline
IllustrisTNG is a set of gravo-magnetohydrodynamical simulations based on the Illustris project
\citep{nelson2018,pillepich2018,naiman2018,marinacci2018,springel2018}. TNG (The Next Generation) has a standard configuration for three different volumes: 35, 75, and 205 cMpc/$h$ of side length -TNG50, TNG100, and TNG300, respectively-. The first box size involves the largest resolution, instead of TNG300, which covers a more vast cosmological region at expense of reducing the gravitational softening. Each simulated box has different levels of resolution (moving from 1 to 3-4, with decreasing numerical resolution and lighter simulation outputs). The assumed cosmology is Planck 2015 with $h =$ 0.6774. In particular, for the comparison intended in this work, we only consider TNG100-1 run, with 2$\times$ 1820$^{3}$ particles, with average cell masses of 7.5  $\times$ 10$^{6}$M$_{\odot}$ and 1.4 $\times$ 10$^{6}$M$_{\odot}$ for dark matter and gas, respectively.
\item \textsc{Flares}\newline
First Light and Reionisation Epoch Simulations (\textsc{Flares}) are a suite of zoom simulations that focuses on the typical overdensities reached during the Epoch of Reionization. The models are presented and discussed in \citet{lovell2021,vijayan2021}, and they are a re-simulated version of \textsc{Eagle} with a total volume of (3.2 cGpc)$^{3}$ -dark matter only-. The dark matter particles have a mass of 8.01$\times$ 10$^{10}$M$_{\odot}$/$h$. Smaller regions of 15 Mpc/$h$ in radius are re-compute with full hydrodynamical treatment (about our boxes in size) from $z =$ 10 down to $z =$ 4.67 with the \textsc{Eagle} galaxy formation scheme. \textsc{Flares} assumes a Planck 2014 cosmological parameters with $h =$ 0.6777, and the same configuration as the \textsc{Eagle} reference run (100 cMpc) with 9.7 and 1.6 $\times$ 10$^{6}$M$_{\odot}$ initial masses for dark matter and gas particles, respectively, leading to a numerical resolution of 2.66 ckpc (between the gravitational softening reached by our synthetic boxes with 18 and 25 cMpc/$h$).
\end{itemize}

\section{Galaxy properties in our simulations}
\label{sec:3}

Following the assumptions from previous section, AGN feedback is not implemented in our simulations \citep{tescari2014}. The simulations were anchored at $z =$ 8, and observables as the cosmic star formation rate and the galaxy stellar mass function were used to calibrate the mass loading factor for the winds $v_{\text{fid}} =$ 600 km/s, and match the observations at the time when \citet{garcia2017b} was published. Hence, the SFR ($z =$ 8) and the stellar mass function are not predictions of the model.\newline

\subsection{Halo occupation fraction}

\begin{figure}
\centering
\includegraphics[scale=0.4]{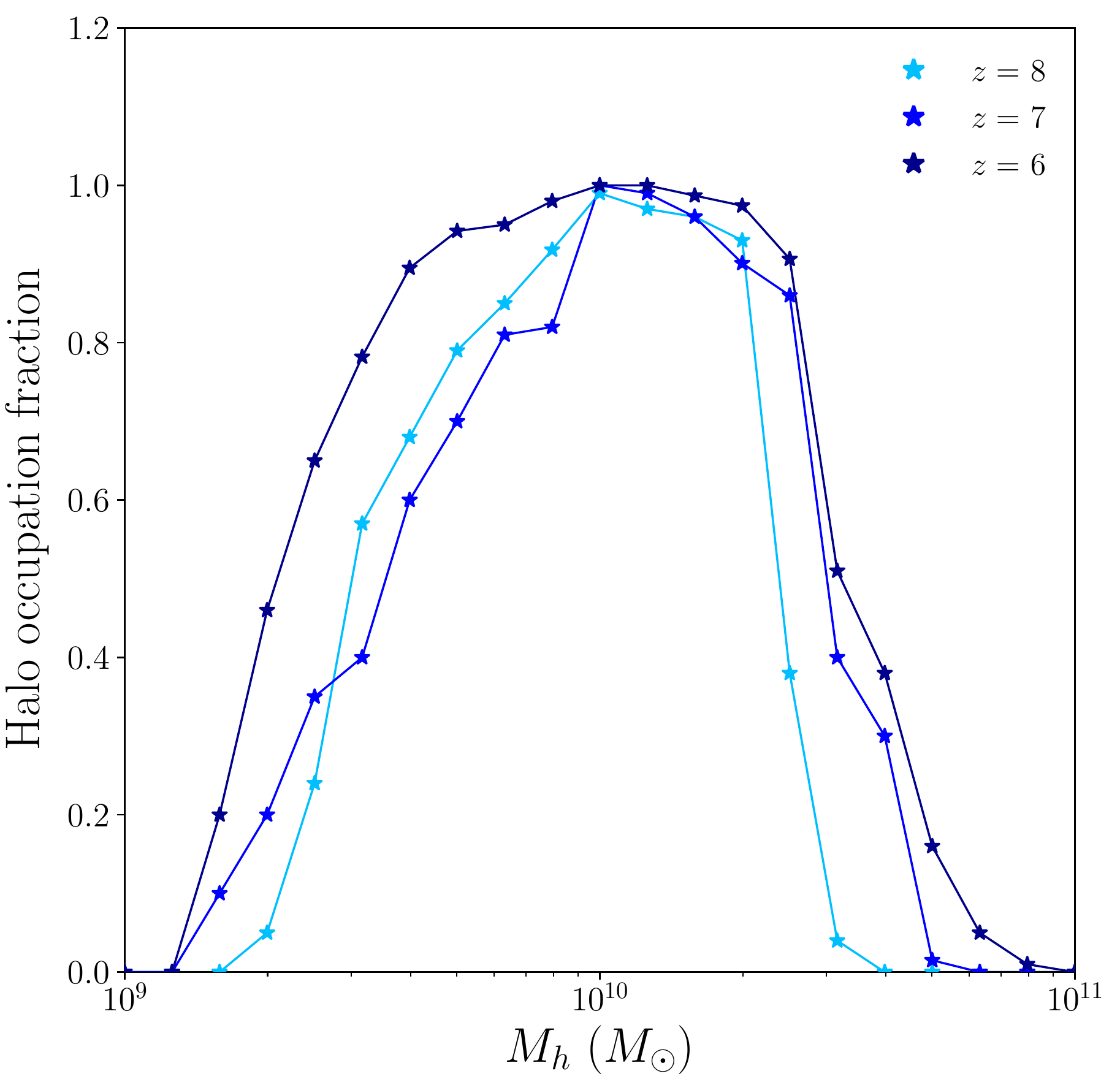}
\caption{\small{Halo occupation fraction in the simulation Ch 18 512 MDW (our fiducial run). The distributions show the percentage of dark matter halos containing formed stars at $z =$ 8, 7, and 6 (light, navy, and dark blue lines, respectively).}}
\label{fig:7}
\end{figure}

The halo occupation fraction presents the distribution of dark matter halos with chemical enriched star-particles at a particular redshift, as a function of the halo mass.
We present the halo occupation fraction in the fiducial model Ch 18 512 MDW in Fig.~\ref{fig:7}, including mass bins of log$(M_h/M_{\odot})=$ 0.1. It is worth noting that we only take into account halos above the mass resolution $M_{h, min}=$ 1.48 $\times$ 10$^{9}M_{\odot}$.\newline

The distribution in Fig.~\ref{fig:7} reaches its maximum occupation at $M_h \sim$ 1 $\sim$ 10$^{10}M_{\odot}$ for $z =$ 8, 7, and 6, which is consistent with the hierarchical model of the structure formation. Dark matter halos with masses below 10$^{9}M_{\odot}$ contain less than 470 dark matter particles, then they cannot be considered virialized and, the star formation in such regions is disfavoured for two reasons: first, they are unresolved, and they do not efficiently experience atomic cooling; hence, gas is collisionally excited and less likely to form structure in the dark matter wells until it cools down. On the contrary, halos with large masses ($\geq$ 10$^{10}M_{\odot}$) preferentially form stars, but they are rare on the simulations at high redshift. Although scarce, dark matter halos with large masses present a non-negligible occupation of chemically enriched stellar populations.\newline

At $z =$ 6 (dark blue line), when the Universe has evolved for $\sim$ 360 Myr from the start of the simulations ($z =$ 8), the highest probability of finding fully occupied halos occurs at $M_h \sim$ 1 $\times$ 10$^{10}$M$_{\odot}$. The latter result indicates that halos in the simulation have grown in mass during this period and, consequently, the cosmic star formation rate.

\subsection{Stellar-to-halo mass function}

\begin{figure}
\centering
\includegraphics[scale=0.4]{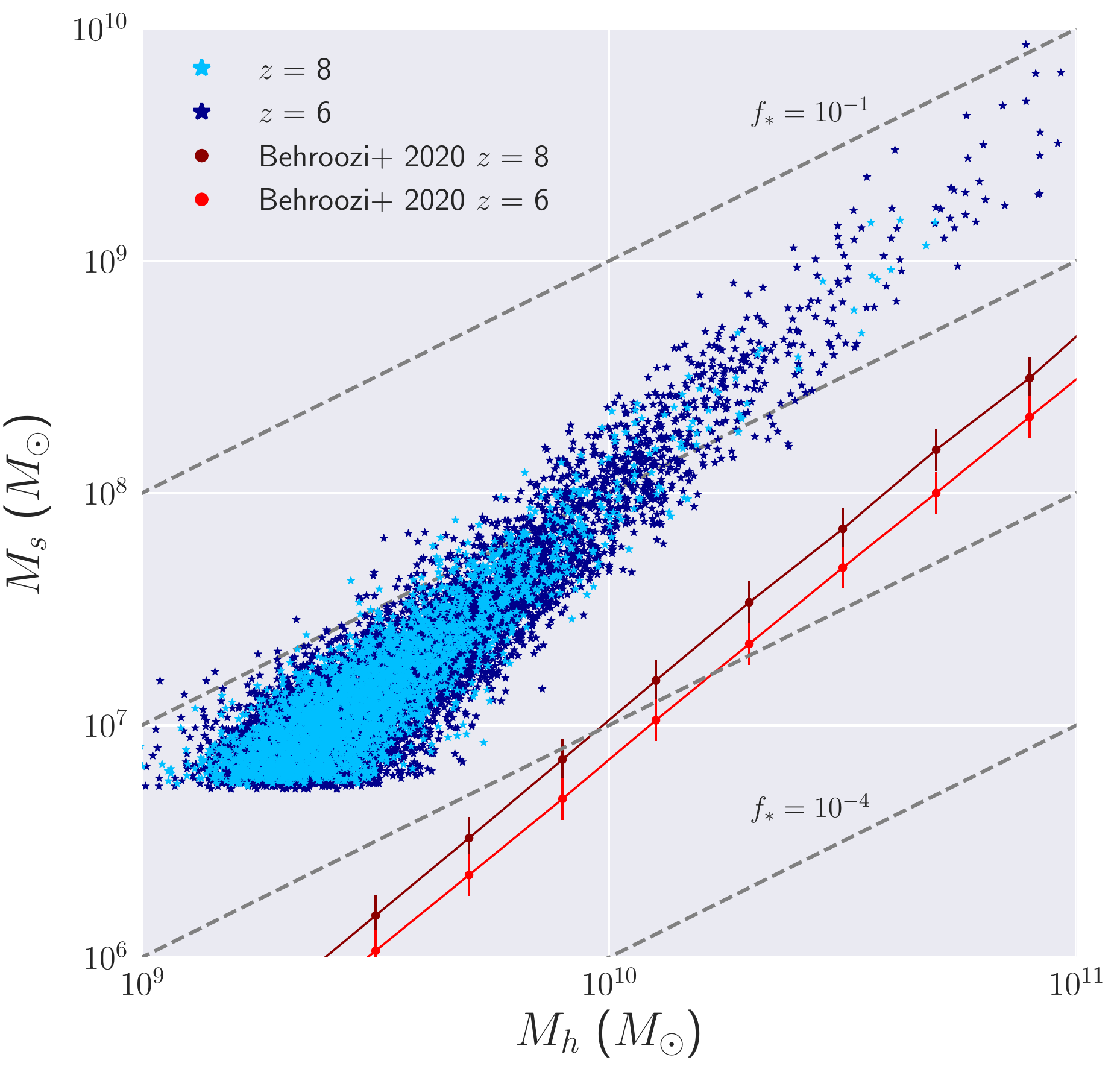}
\caption{\small{Galaxy stellar-to-halo virial mass function in our reference model Ch 18 512 MDW. The graph displays the distribution of masses above the threshold for this simulation box-size at redshifts $z =$ 8 and 6 (corresponding to light and dark blue stars, respectively). For comparison, the predicted trends from \textsc{UniverseMachine} \citep{behroozi2020} are shown at $z =$ 8 (dark red circles) and $z =$ 6 (red dots). Grey dashed lines indicate constant stellar functions $f_* = M_*/M_h$, ranging from 10$^{-4}$ (bottom) to 10$^{-1}$ (top).}}
\label{fig:4}
\end{figure}

Another observable that we check in our models is known as the galaxy stellar-to-halo virial mass function, presented in Fig.~\ref{fig:4} for the fiducial run Ch 18 512 MDW, at $z =$ 8 and 6. \newline

Fig.~\ref{fig:4} presents the distribution of the stellar-to-halo mass for each galaxy in the fiducial simulation above the mass resolution threshold (> 10$^{9}M_{\odot}$ or equivalently, more than 470 dark matter particles). Simulated galaxies follow a trend of 10$^{-2}$, and at the high mass end, $f_*$ moves to 10$^{-1}$. $f_*$ evolution grows with redshift, with more galaxies with large stellar masses at $z =$ 6. This result is also seen in numerical simulations as \textsc{Astraeus} \citep{hutter2020}, with $M_{h} =$ 10$^{9.5-10}M_{\odot}$ corresponding to stellar masses $\sim$ 10$^{7.5-8}M_{\odot}$ during the Epoch of Reionization. Moreover, \citet{oshea2015} display this distribution (Fig. 2, left), but they stack all the galaxies in their realizations up to their final redshifts. The \textsc{Renaissance} simulations describe galaxies that are evolving during the progress of Reionization; therefore, they account for halo masses down to 7 $\times$ 10$^{6}M_{\odot}$ (halos with formed stars). Their plot is not directly comparable with Figure~\ref{fig:4} because the mass range covered by their simulations differs from ours. Still, galaxies around 10$^{9}M_{\odot}$ in their work show the same difference of two orders of magnitude in $M_{h}/M_{s}$. Recent observations by \citet{stefanon2021} reveal the same ratio of two orders of magnitude for stellar-to-halo mass ratios (see Table 7 in their work). They emphasize that there is no significant evolution in the observed stellar-to-halo mass function for galaxies in the first Gyr of cosmic time. The latter conclusion is consistent with Fig.~\ref{fig:4}. Instead, predictions by \citet{behroozi2020} are an order of magnitude below the trend of our fiducial simulation, at $z =$ 8 and 6. However, we cannot provide a clear explanation for this discrepancy since \textsc{UniverseMachine} is calibrated to resolve virialized halos up to $z \sim$ 15, but the ratio predicted by their simulation is off compared with other simulations at $z =$ 6 - 8 (including ours).\newline

It is worth mentioning that we only calculate the stellar-to-halo mass function for the fiducial model Ch 18 512 MDW because there is minimal variation in the range of masses resolved by our simulations due to our small boxes (12, 18, and 25 cMpc/$h$). Thus, the trend shown in Figure~\ref{fig:4} is barely affected by changes in the box size.

\subsection{Galaxy stellar mass function}

One can also count the number density of galaxies formed inside the virialized halos per unit volume $V$ per stellar mass bin $\Delta M$ \citep{weigel2016}. This observable is known as the galaxy stellar mass function, and it is given by:
\begin{equation}
\Phi (z) = \frac{\#_{\text{gal}}(\Delta M)}{V \cdot \Delta M}.
\end{equation}

The analytical form that describes the galaxy stellar mass function is commonly described using a Schechter function \citep{schechter1976}, as follows:\newline
\begin{equation}
\label{gsmf}
\Phi (M) = \text{ln(10)}\Phi^*\text{e}^{-10(M - M_{*})}\text{10}^{(M - M_{*})(\alpha + 1)}.
\end{equation}
\noindent The exponential term in the expression above shows the evolution for the high- and low-mass and the fore-most right term a power-law behavior as a function of the stellar mass.\newline

We present the best fit parameters and corresponding errors for the galaxy stellar mass functions at $z =$ 8, 7, and 6 for three of our models with same configuration (MDW and no molecular cooling) with box sizes of 12, 18, and 25 cMpc/$h$, in Table~\ref{table_gsmf}. We derive this parameters with an adapted version from python routine \textsc{emcee}. Table~\ref{table_gsmf} also shows the Schechter function parameters from \citet{duncan2014}, \citet{grazian2015}, \citet{rachana2019} and \citet{stefanon2021}.\newline

One highlight from the Table~\ref{table_gsmf} is that the slope of the galaxy stellar mass function at $z =$ 6 - 8 remains constant, and in all the cases presented, is close to the value -2. These findings are consistent with the observational constraints also shown in Table~\ref{table_gsmf}.\newline

\begin{table}
\centering
\caption{\footnotesize{Best fit Schechter function parameters and uncertainties for the galaxy stellar mass function $\Phi (M)$. For each redshift, we find the best parameters for the fiducial model and two equivalent runs with the same setup, but 12 and 25 cMpc/$h$ box side.}}
\vspace{0.5cm}
\resizebox{0.8\textwidth}{!}{%
\begin{tabular}{|lccc|} 
  \hline
  & & & \\
  & log$_{\text{10}}M_{*}$ & $\alpha$ & $\Phi^*$ (10$^{-5}$Mpc$^{-3}$)  \\ 
  & & & \\ \hline 
$z \sim$ 8  & & & \\\hline
\citet{stefanon2021} & 9.98$^{+0.44}_{-0.24}$  & -1.82$^{+0.20}_{-0.21}$  & 2.04$^{+0.35}_{-0.78}$  \\ 
\citet{rachana2019} & 10.54$^{+1.00}_{-0.94}$  & -2.30$^{+0.51}_{-0.46}$  & 0.095$^{+0.56}_{-0.08}$ \\ 
Ch 18 512 MDW  & 10.34$\pm$0.02 & -2.20$\pm$0.05 & 0.092$\pm$0.005\\ 
Ch 12 512 MDW  & 9.25$\pm$0.02 & -2.15$\pm$0.06 & 0.870$\pm$0.005\\ 
Ch 25 512 MDW  & 10.55$\pm$0.02 & -2.30$\pm$0.07 & 0.098$\pm$0.005\\ 
\hline
$z \sim$ 7 &  & & \\ \hline
\citet{stefanon2021} & 10.04$^{+0.15}_{-0.13}$  & -1.73$^{+0.08}_{-0.08}$  & 7.24$^{+0.62}_{-0.71}$ \\ 
\citet{rachana2019} & 10.27$^{+0.60}_{-0.67}$  & -2.01$^{+0.17}_{-0.13}$  & 3.9$^{+9.2}_{-2.85}$  \\ 
\citet{grazian2015} & 10.69$^{+1.58}_{-1.58}$  & -1.88$^{+0.36}_{-0.36}$  & 0.57$^{+59.68}_{-0.56}$ \\ 
\citet{duncan2014}  & 10.51$^{+0.36}_{-0.32}$  & -1.89$^{+1.39}_{-0.61}$  & 3.60$^{+3.01}_{-0.35}$  \\ 
Ch 18 512 MDW  & 10.61$\pm$0.02 & -1.95$\pm$0.04 & 0.67$\pm$0.03\\ 
Ch 12 512 MDW  & 10.72$\pm$0.02 & -1.92$\pm$0.04 & 0.60$\pm$0.03\\ 
Ch 25 512 MDW  & 10.51$\pm$0.02 & -2.10$\pm$0.05 & 0.72$\pm$0.03\\ 
\hline
 $z \sim$ 6 &  & & \\  \hline
\citet{stefanon2021} & 10.24$^{+0.08}_{-0.11}$  & -1.88$^{+0.06}_{-0.03}$  & 8.13$^{+0.52}_{-0.35}$\\ 
\citet{rachana2019} & 10.35$^{+0.50}_{-0.50}$  & -1.98$^{+0.07}_{-0.07}$  & 6.05$^{+8.96}_{-3.49}$   \\ 
\citet{grazian2015} & 10.49$^{+0.32}_{-0.32}$  & -1.55$^{+0.19}_{-0.19}$  & 6.91$^{+13.5}_{-4.57}$  \\ 
\citet{duncan2014}  & 10.87$^{+1.13}_{-0.54}$  & -2.00$^{+0.57}_{-0.40}$  & 1.4$^{+41.1}_{-1.4}$  \\ 
Ch 18 512 MDW  & 10.41$\pm$0.02 & -2.01$\pm$0.05 & 1.20$\pm$0.03\\ 
Ch 12 512 MDW  & 10.42$\pm$0.02 & -2.02$\pm$0.05 & 1.15$\pm$0.03 \\ 
Ch 25 512 MDW  & 10.40$\pm$0.02 & -2.05$\pm$0.05 & 3.20$\pm$0.03\\ 
\hline
\end{tabular}
\label{table_gsmf}
}
\end{table}

\begin{figure}
\centering
\includegraphics[scale=0.43]{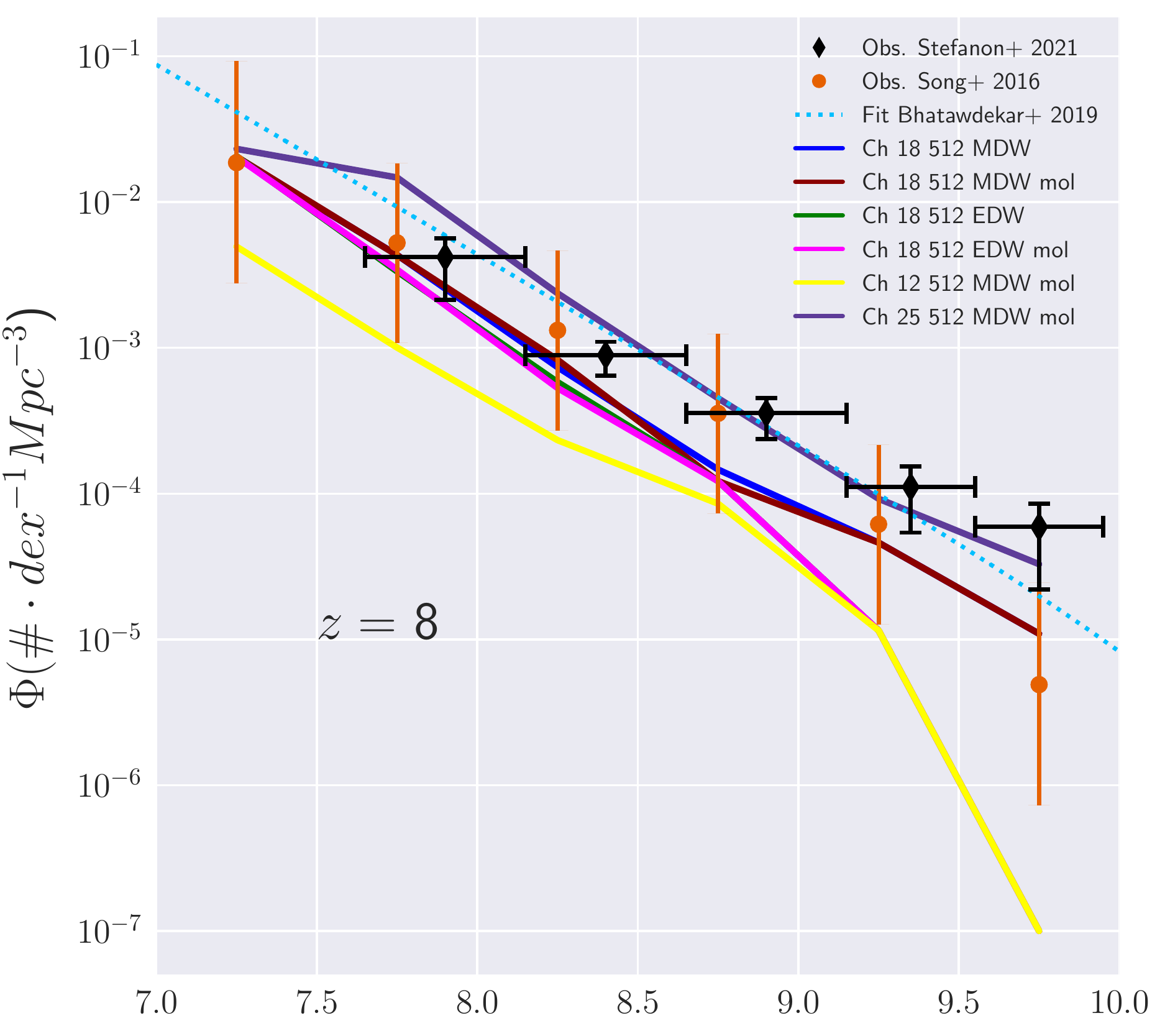}\\
\includegraphics[scale=0.43]{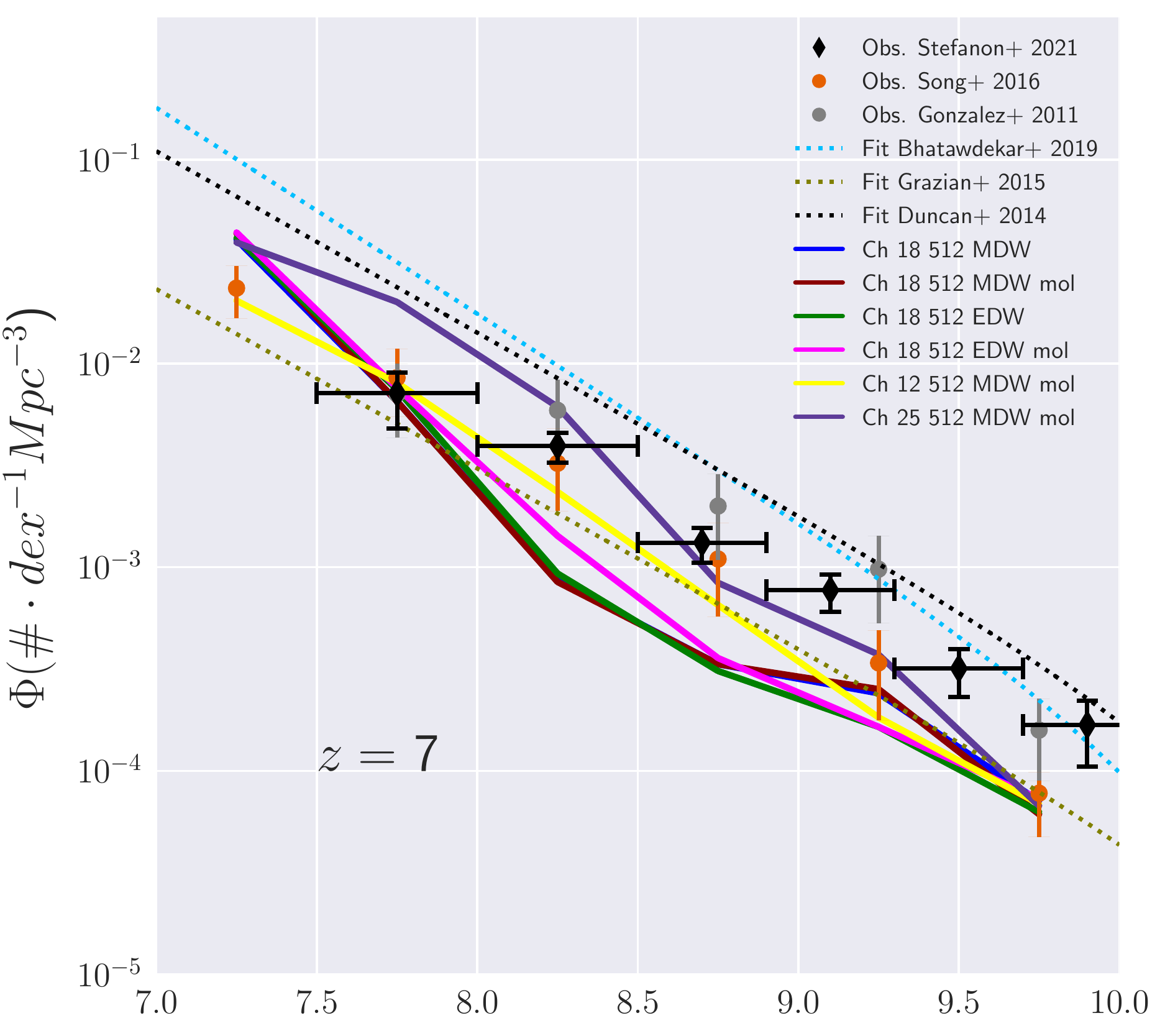}\\
\includegraphics[scale=0.43]{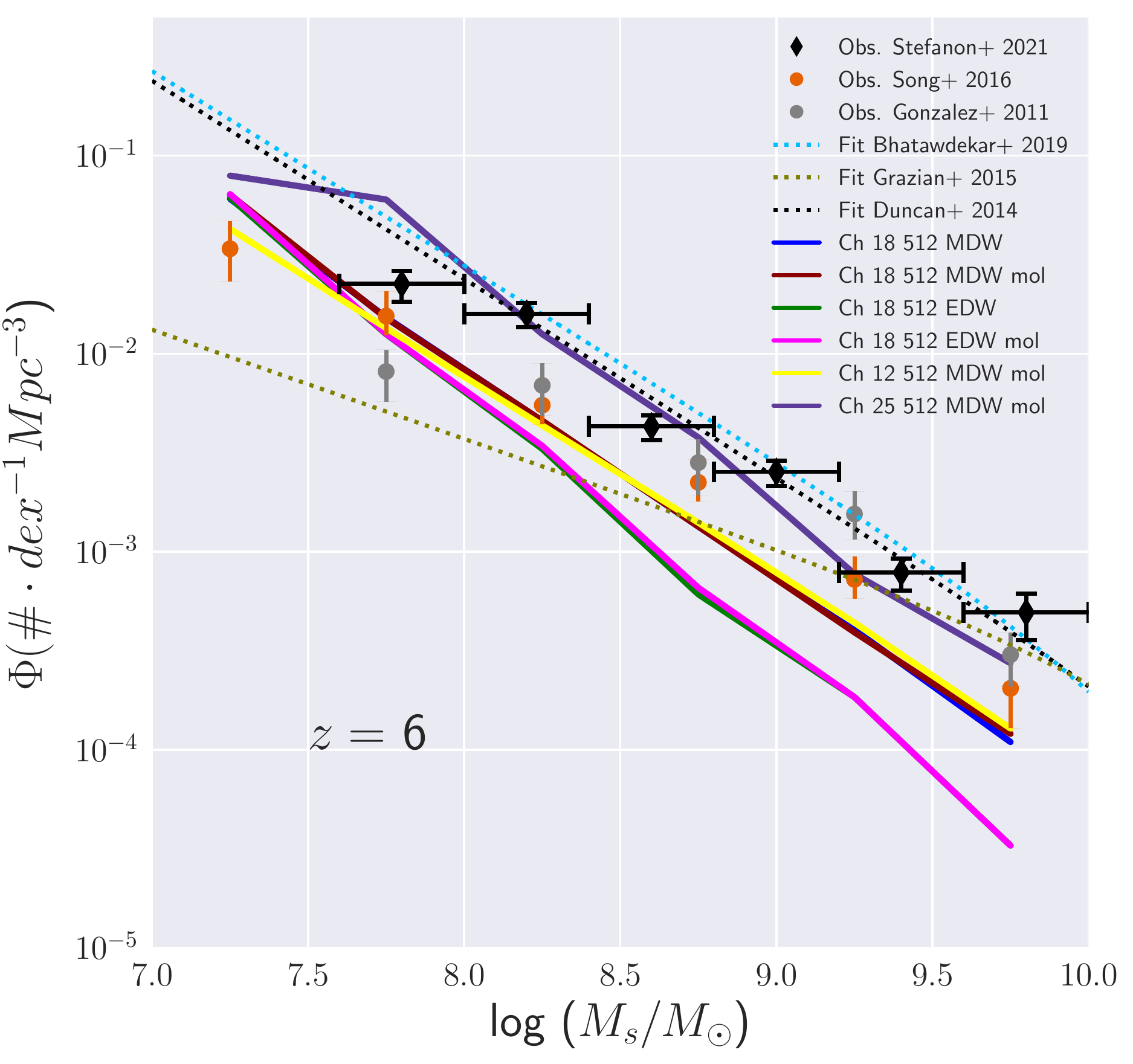}\\
\caption{\small{Simulated galaxy stellar mass function at $z =$ 8, 7 and 6 in the top, middle and bottom panels. Theoretical predictions from our models are compared with observations by \citet{stefanon2021} in black diamonds, \citet{song2016} in orange circles and \citet{gonzalez2011} in grey circles, and the best fits proposed \citet{rachana2019}, \citet{grazian2015} and \citet{duncan2014} in light blue, olive and black dotted lines, respectively. The parameters of the Schechter function~\eqref{gsmf} for each model are presented in Table~\ref{table_gsmf}.}}
\label{fig:5}
\end{figure}

The evolution of the galaxy stellar mass function from $z = $ 8 to 6 is displayed in Fig.~\ref{fig:5}. Remarkably, the galaxy stellar mass function $\Phi$ at $z =$ 8 was set to match observations by \citet{song2016} and calibrate the mass loading factor of the simulations $v_{\text{fid}}$. However, the theoretical trends agree well with the best fit at $z =$ 8 by \citet{rachana2019} and with the most recent Spitzer/IRAC observations by \citet{stefanon2021}, which is quite reassuring since the observational detections came afterward than our simulations. \newline

The predicted galaxy stellar mass function at $z = $ 8, 7, and 6 are compatible with the observational data at high $z$. Nonetheless, the simulations slightly differ from the galaxy stellar mass function reported by \citet{song2016} and \citet{gonzalez2011} in the high mass end at $z =$ 6, mainly because massive galactic halos are scarce in the simulations at these redshifts; thus, high-mass galaxies are rare. Larger synthetic boxes could alleviate this mass bias (for instance, \textsc{Flares} or \textsc{UniverseMachine}). \newline

On the other hand, results with \textsc{Flares} \citep[][supported by a vast cosmological box, that at $z =$ 5 and 6, reach stellar masses up to 10$^{11}$ M$_{\sun}$]{lovell2021,vijayan2021}, claim that the galaxy stellar mass function must be fitted with a double-slope Schechter function with a knee at M$_{s} =$ 10$^{10}$M$_{\odot}$. They back the latter argument with recent observational constrains from \citet{stefanon2021}. Nevertheless, our small boxes do not allow us to test this range of mass (M$_{s,\text{max}} \sim$ 10$^{9.8}$M$_{\odot}$ in our largest realization; hence, we keep a single-slope fit). One of the motivations for \citet{lovell2021} work was to extend the range of stellar masses and the number of resolved galaxies reached by the \textsc{Eagle} simulation. Our cosmological runs have similar modules, box sizes, and an analogous configuration as \textsc{Eagle}; therefore, tests related to galaxy observables must be done with \textsc{Eagle} -or equivalent hydrodynamical simulations-. Instead, this particular conclusion derived by \citet{lovell2021} is out of the reach of our simulations.

\subsection{Halo mass function}

\begin{figure}
\centering
\includegraphics[scale=0.4]{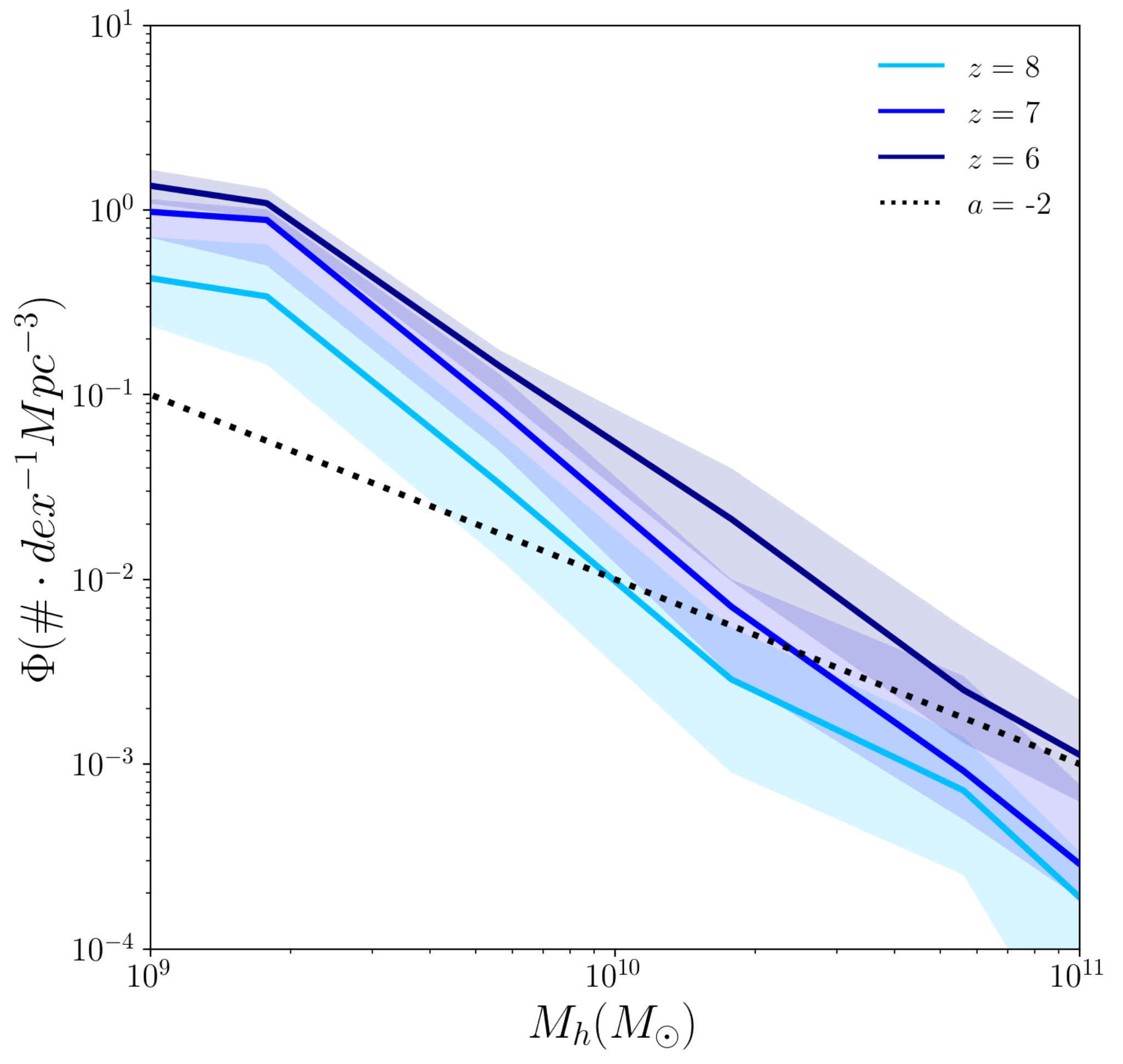}
\caption{\small{Halo mass functions at $z =$ 6, 7, and 8 in dark, navy, and light blue. We include shadow regions corresponding to Poisson errors for the  fiducial run Ch 18 512 MDW. As a reference, we show a constant power-law slope $\alpha =$ -2 for galaxies at high redshift, in a dotted black line.}}
\label{fig:6}
\end{figure}

In order to characterize the galaxies in the simulations, the halo mass functions at redshifts $z =$ 8, 7, and 6 are presented in Fig.~\ref{fig:6}, in light, navy, and dark blue lines, respectively. This quantity is computed with systems above the mass resolution limit (a resolved halo in the simulation contains $\sim$ 470 dark matter particles, or equivalently, a minimum mass $M_{h, min}=$ 1.48 $\times$ 10$^{9}M_{\odot}$ for boxes of 18 Mpc/$h$ length side). Whenever a galactic halo is below this mass threshold, the object is considered unresolved and is excluded from the statistics.\newline
The evolution of the mass function for dark matter halos is computed only with the fiducial model Ch 18 512 MDW since the number of resolved galaxies is almost independent of the feedback mechanisms or the cooling processes implemented in the simulations. As a reference, a dotted black line is presented on top of our predictions in Fig.~\ref{fig:6}, indicating a constant faint-end slope $\alpha =$ -2.\newline

Theoretically, the number density of halos follows the relations $\frac{d(log N)}{d(log M_h)}=$ -1, as shown in Fig.~\ref{fig:6}, and $N \propto M_h^{\alpha + 1}$, leading to a faint-end slope $\alpha =$ -2\footnote{See a similar discussion in \citet{behroozi2020}.}. Although the set of simulations presented in this work are unable to provide a direct prediction on the power slope due to the narrow range of halo masses and the small box sizes of the simulations, the curves in Fig.~\ref{fig:6} show a trend consistent with a power-law slope of -2 at $z \sim$ 6-8.\newline

We support the latter claim based on the results described in previous sections. The stellar-to-halo mass ratio (Fig.~\ref{fig:4}) shows little evolution of the mass ratio at $z =$ 8 to 6 (regardless of the increasing number of halos that form galaxies with time). On the other hand, the galaxy stellar mass function (Table~\ref{table_gsmf} and Fig.~\ref{fig:5}) indicates that the slope is close to -2 during the time frame described by the simulations. Since the halo masses, $M_{h}$ are two orders of magnitude larger than the stellar masses $M_{s}$ -this ratio stays constant at the tail of the Reionization- and the slope for the stellar mass function is -2, with almost no variation in time, the value of the slope of the halo mass function is consistent with -2.

\subsection{Star formation rate density}

The star formation rate is the mass of the new stars in the simulation, measured in the total volume per year. It is commonly assessed by galaxy surveys or derived from studies with luminosity functions. There are two ways to compute the cosmic star formation rate in the numerical runs: i) adding up the star formation of each gas particle, per comoving volume $V$; or ii) recovering the SFR estimate for galaxy groups from the FoF catalog.\newline

\begin{figure}
\centering
\includegraphics[scale=0.45]{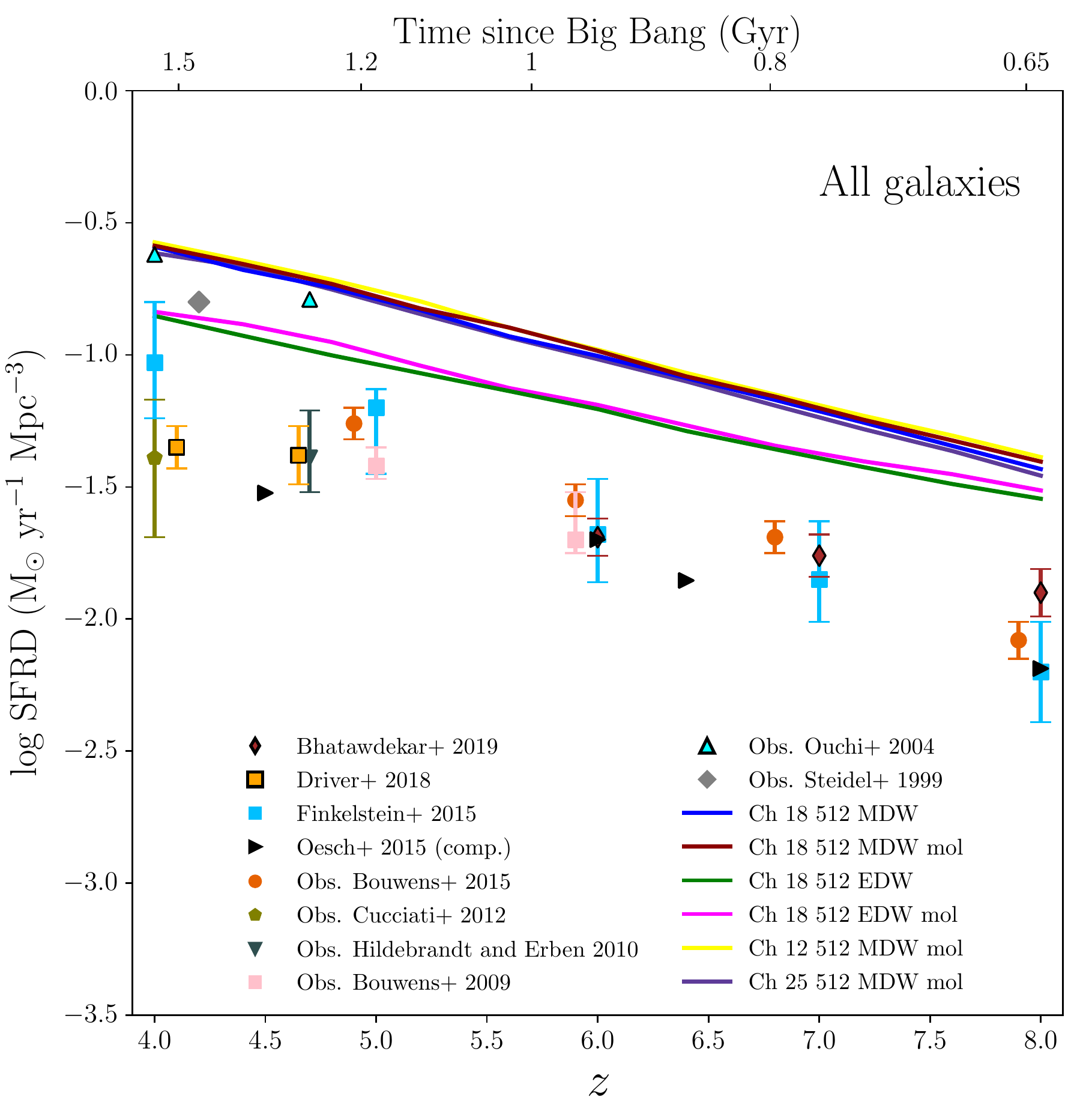}
\includegraphics[scale=0.45]{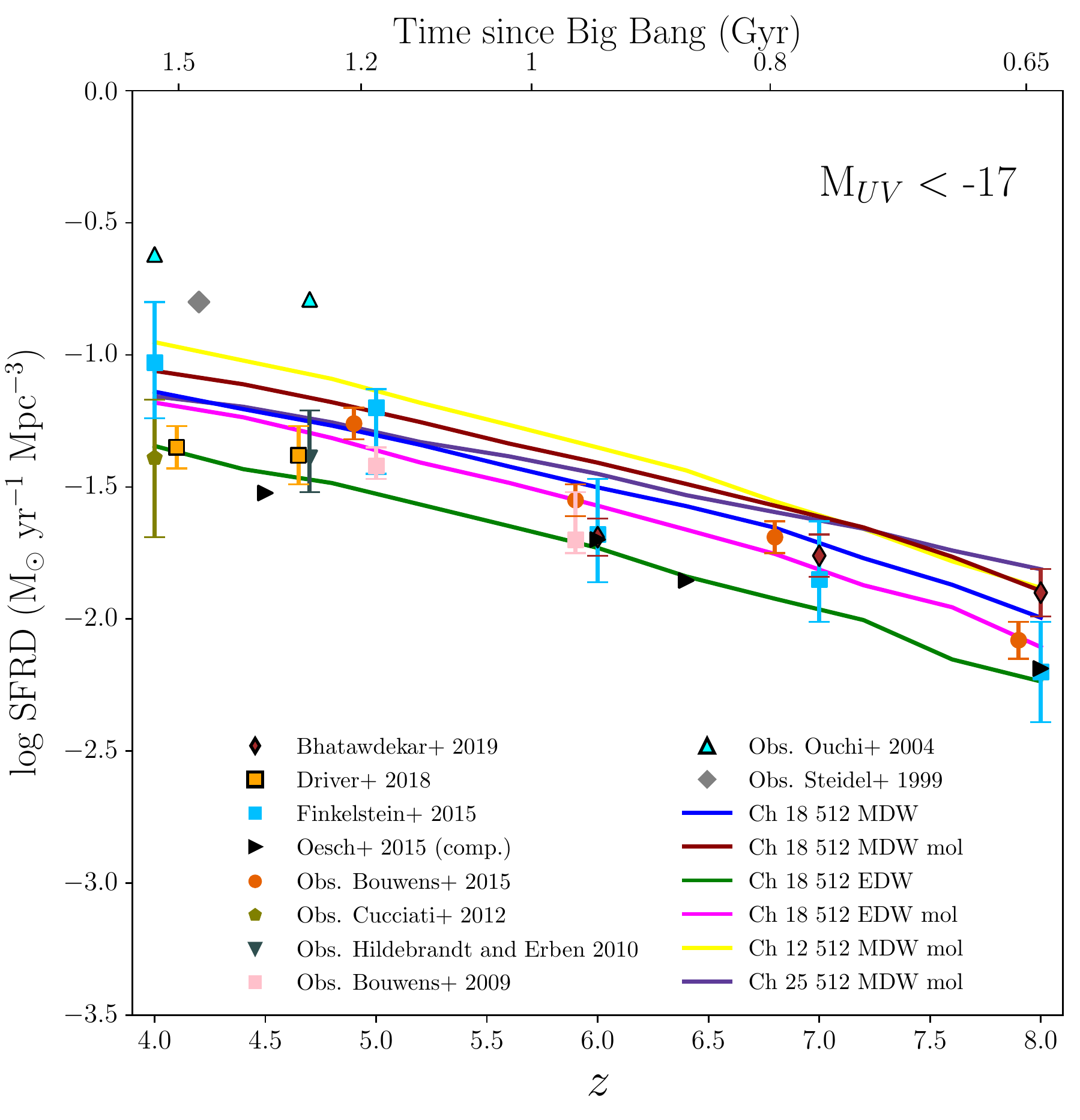}
\caption{\small{Cosmic star formation rate density in the redshift range of 4 $< z <$ 8. The predictions from the simulations are compared with observations from \citet{rachana2019} in brown diamonts, \citet{driver2018} in yellow squares, \citet{bouwens2015} in orange circles (with dust corrections), \citet{cucciati2012} in olive pentagon, \citet{hildebrandt2010} in green inverted triangle, \citet{bouwens2009} in pink square, \citet{ouchi2004} in cyan triangles and \citet{steidel1999} in grey diamond. In the left panel, the SFR in the simulations is computed including all objects in the box per unit volume. On the right, the observable is limited to masses with the luminosity cut of M$_{\text{UV}} <$ -17, equivalent to a minimum SFR of $>$ 0.331 $M_{\sun}/$yr, following the observational constraints of our current telescopes.}}
\label{fig:sfr_thrs}
\end{figure}

Fig.~\ref{fig:sfr_thrs} shows the cosmic star formation rate in our simulations at 4 $< z <$ 8. The left panel shows the total SFR (including contributions from all the collapsed objects inside the box). Conversely, the right panel displays the same observable, but this time, applying a cut in mass; thus, only the most luminous galaxies are taken into account in the calculation. This mass threshold responds to the resolution achieved by our telescopes that only detects the most luminous galaxies (in particular, at high redshift). Current instruments do not detect the faintest objects; therefore, their SFR cannot be inferred with that method. Consequently, there is an excess in the star formation rate predicted by the simulations -on the left- to observational data. The discrepancy between the calculated and the observed SFR is corrected in the right panel by imposing a luminosity cut M$_{\text{UV}} <$ -17 (corresponding to a minimum SFR $>$ 0.331 $M_{\sun}/$yr and the absolute magnitude set by Hubble observations). When we impose the latter criterium to the simulations, the predicted SFR agrees well with the observations to-date (except for data points measured by \citet{steidel1999} and \citet{ouchi2004} that do not account for dust corrections).\newline

Interestingly, data from \citet{driver2018} and \citet{rachana2019} had not been published when the simulations were run, but most of the models are in agreement with these observations. \newline
On the other hand, the cosmic SFR reported by \citet{finkelstein2016} -with a corresponding comparison with \citet{madau2014}- shows an increment of 1 dex in their reference model, consistent with findings from this work with the mass cut M$_{\text{UV}} <$ -17 (right panel of Figure~\ref{fig:sfr_thrs}).\newline

\begin{figure}
\centering
\includegraphics[scale=0.6]{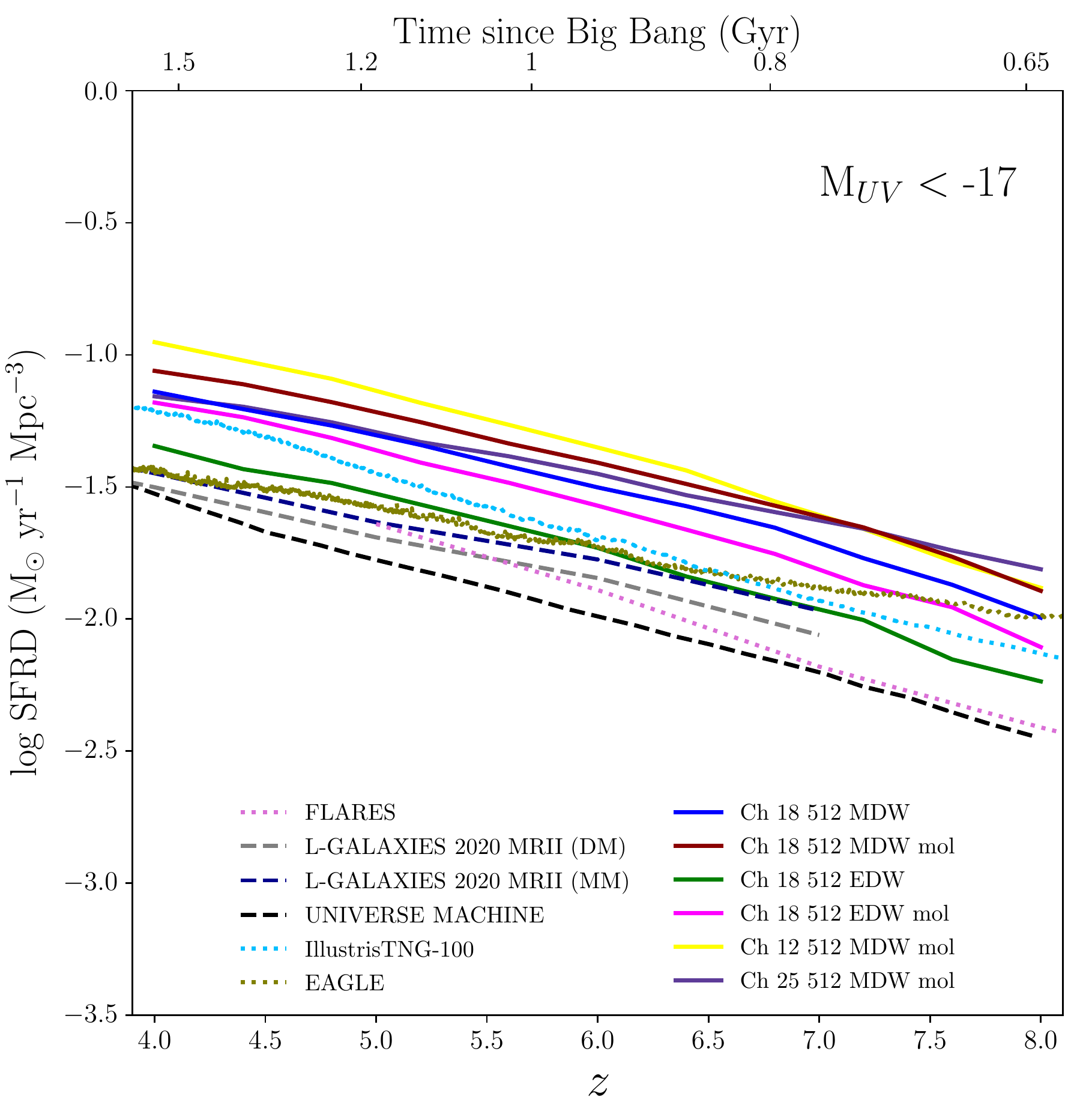}
\caption{\small{Cosmic star formation rate density in the redshift range of 4 $< z <$ 8. The predictions from our models are compared with the calculated SFR with the \textsc{UniverseMachine} \citep{behroozi2020}, L-Galaxies 2020 \citep{henriques2020,yates2021a}, \textsc{Flares} \citep{lovell2021,vijayan2021}, TNG100 \citep{nelson2018,pillepich2018,naiman2018,marinacci2018,springel2018}, and the \textsc{Eagle} simulations \citep{crain2015,schaye2015}. The observable is limited to observed galaxies with absolute magnitude cut of M$_{\text{UV}} <$ -17, equivalent to a minimum SFR of $>$ 0.331 $M_{\sun}/$yr.}}
\label{fig:sfr_sims}
\end{figure}

Figure~\ref{fig:sfr_sims} shows a compilation of theoretical predictions for the cosmic star formation rate by \textsc{UniverseMachine} \citep{behroozi2020}, L-Galaxies 2020 \citep{henriques2020,yates2021a}, \textsc{Flares} \citep{lovell2021,vijayan2021}, TNG100 \citep{nelson2018,pillepich2018,naiman2018,marinacci2018,springel2018}, \textsc{Eagle} \citep{crain2015,schaye2015} and our models.\newline

Except for Ch 18 512 EDW, all our simulations show an excess on the SFR when compared with other theoretical models. As mentioned in a previous section, the simulation with the closest configuration to ours is \textsc{Eagle}, and this explains why their calculated SFR is compatible with our runs with the energy-driven winds prescription for the supernova outflows. This is also true with TNG100, which agrees well with EDW runs, but it is always below the prediction with MDW realizations. This outcome from our simulations is promising since the winds implemented in the IllustrisTNG project have much more complex dynamics than ours: the velocity of the galactic winds $v_w$ also depends on $z$  \citep[suppressing the efficiency of winds at with the Hubble factor, ][]{pillepich2018}, but scales in the same way as our winds with the virial halo mass. Another remarkable difference in the IllustrisTNG winds is that the outflow mass loading is a non-monotonic function of the galaxy stellar mass \citep{nelson2018}. We do not count for such dependence in our models. Finally, TNG introduces an improved mechanism for the AGN feedback, even for a low accretion rate, whereas this work does not account for AGN feedback.\newline
A slightly different scenario is drawn with the two versions of L-Galaxies 2020. Both configurations match the observed SFR at low and intermediate redshifts because their semi-analytical models were built to follow the chemical enrichment at late times, not during Reionization. Besides, their models heavily rely on observations from Damped Lyman systems (DLAs) that are not possible to extend at the redshifts of the EoR \citep{garcia2017b}.
The gap in the calculated SFR grows among our runs and large cosmological simulations, particularly towards $z \rightarrow$ 8. The \textsc{UniverseMachine} and \textsc{Flares} are theoretical models aim to correct the UV luminosity function and provide a forecast for future wide-field surveys, as the Nancy Roman (previously known as \textsc{Wfirst}), \textsc{Euclid} or JWST. Findings from these large volume boxes, with broader redshift ranges, are poorly constrained by periodic hydrodynamical simulations due to their limited volume, and consequently, reduced number of massive galaxies. \newline

It is worth mentioning that the different supernova feedback mechanisms play a dominant role in the evolution of the star formation rate. Fig.~\ref{fig:sfr_thrs} and ~\ref{fig:sfr_sims} show lower values for the SFR with energy-driven winds than momentum-driven winds (EDW and MDW, respectively), indicating that the former mechanism is more efficient at quenching the SFR because it prevents the overcooling present in the latter case that leads to excess on the number of stars that would form during a time interval of $\sim$ 1 billion years ($z =$ 8 to 4). Notably, once many star formation events occur in the simulation, the stochastic SFR converges to its continuous history, and the galaxies grow in size and mass through this physical scheme.\newline

Finally, it is interesting to study the ratio between the observed $M_{\text{UV}} <$ -17 and total cosmic star formation rates in the different realizations considered in Fig.~\ref{fig:sfr_thrs}. Although, this is not an observable, it reflects how the mass cut affects the overall SFR in the synthetic realizations.\newline
\begin{figure}
\centering
\includegraphics[scale=0.47]{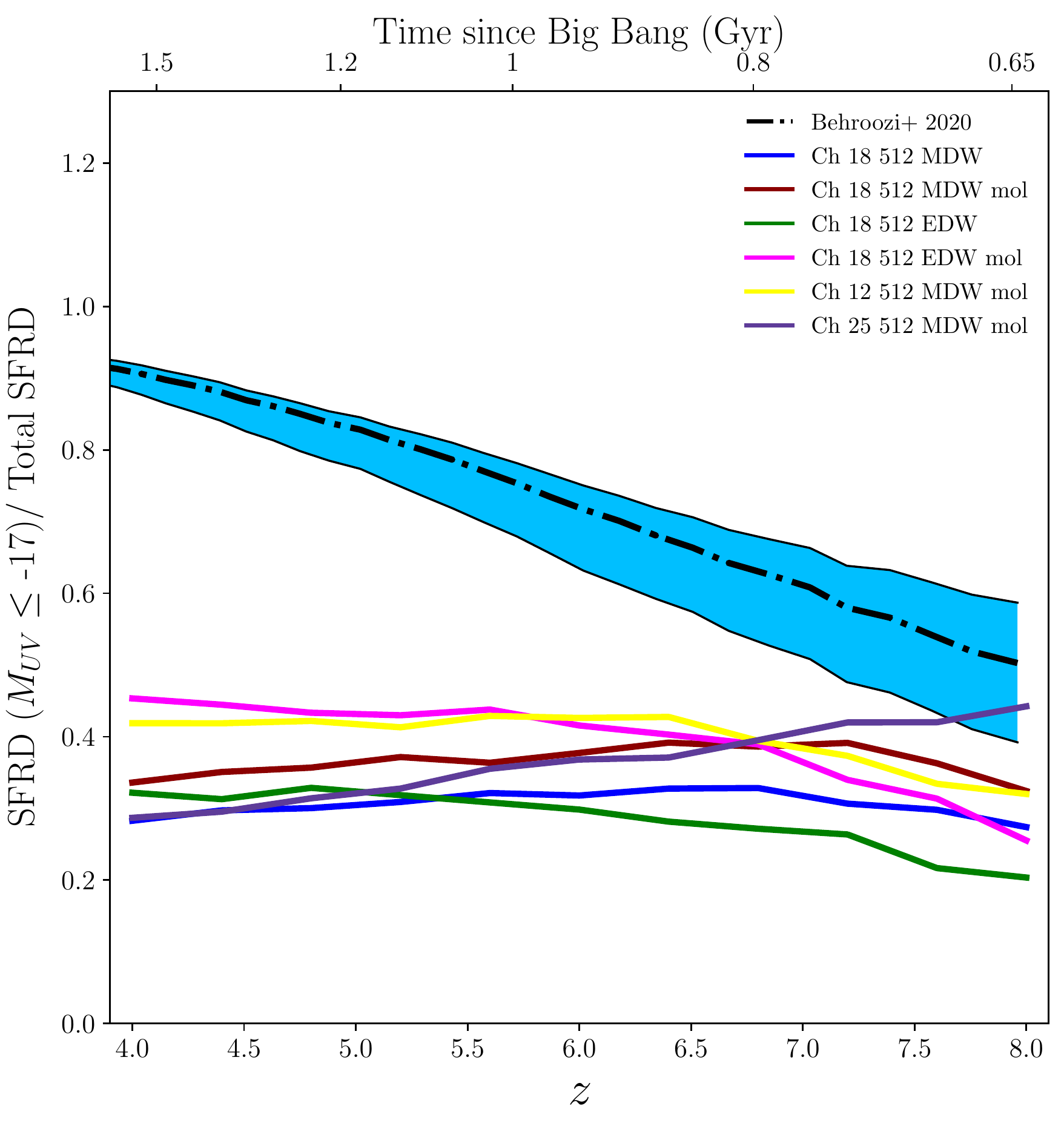}
\caption{\small{Predicted ratio between the observed $M_{\text{UV}} <$ -17 and total cosmic star formation rates from the simulations, and comparison with \textsc{UniverseMachine} ratio. The blue band indicates the error, and the dashed black line the mean value from that set of simulations.}}
\label{fig:3}
\end{figure}

Fig.~\ref{fig:3} shows the predicted ratio from the simulations and \textsc{UniverseMachine}. Most of our configurations show a flat trend at all the redshift range (except for the Ch 25 512 MDW mol, which is, in fact, the simulation with the lowest resolution). This result leads to the conclusion that  $\sim$ 2 out of 5 simulated galaxies are about the luminosity cut of $M_{\text{UV}} <$ -17, and this ratio does not evolve from $z =$ 8 to 4. Conversely, \citet{behroozi2020} show a rapid increment in the SFRD($M_{\text{UV}} <$ -17) / Total SFRD, from 0.5 to 0.9 for $z =$ 8 to 4, respectively. The latter is consistent with a change of 1 dex in Fig.~\ref{fig:sfr_thrs} -right panel-, indicating the vast majority of the stars formed during a time frame of 900 Myr are above the luminosity cut M$_{\text{UV}} <$ -17. Beyond the percentage inferred from our simulations, observations show an increasing number of bright galaxies at the tail of Reionization and indirectly confirm the predicted ratio by \citet{behroozi2020}.

\subsection{Chemical enrichment in the simulations}

One of the strengths of this model is the self-consistent chemical enrichment implementation. The metals' production, spread, and mixing to the cincum- and intergalactic medium come from the assumed stellar lifetime function, stellar yields, and initial mass function.\newline

It is worth noting that there are no measurements of the cosmic mass densities for any metal. The only tight constraint is that elements except for H and He should account for about 1\% of the baryonic content in the Universe. This issue becomes even more challenging at high redshift when indirect methods are less precise to quantify the amount of any element.  However, astronomers can estimate lower limits for the percentage of individual metals in the total census by measuring the total mass density for metal ions in the IGM (see work from \citet{garcia2017b} with CII and CIV, and the detections by \citet{codoreanu2018} on SiII and SiIV). Also, it is possible an approximative evaluation of the relative metallicity to Hydrogen with damped Ly$\alpha$ systems, but this method does not provide any observational constraints for O or Si (among other metals).\newline

\begin{figure}
\centering
	\includegraphics[scale=0.45]{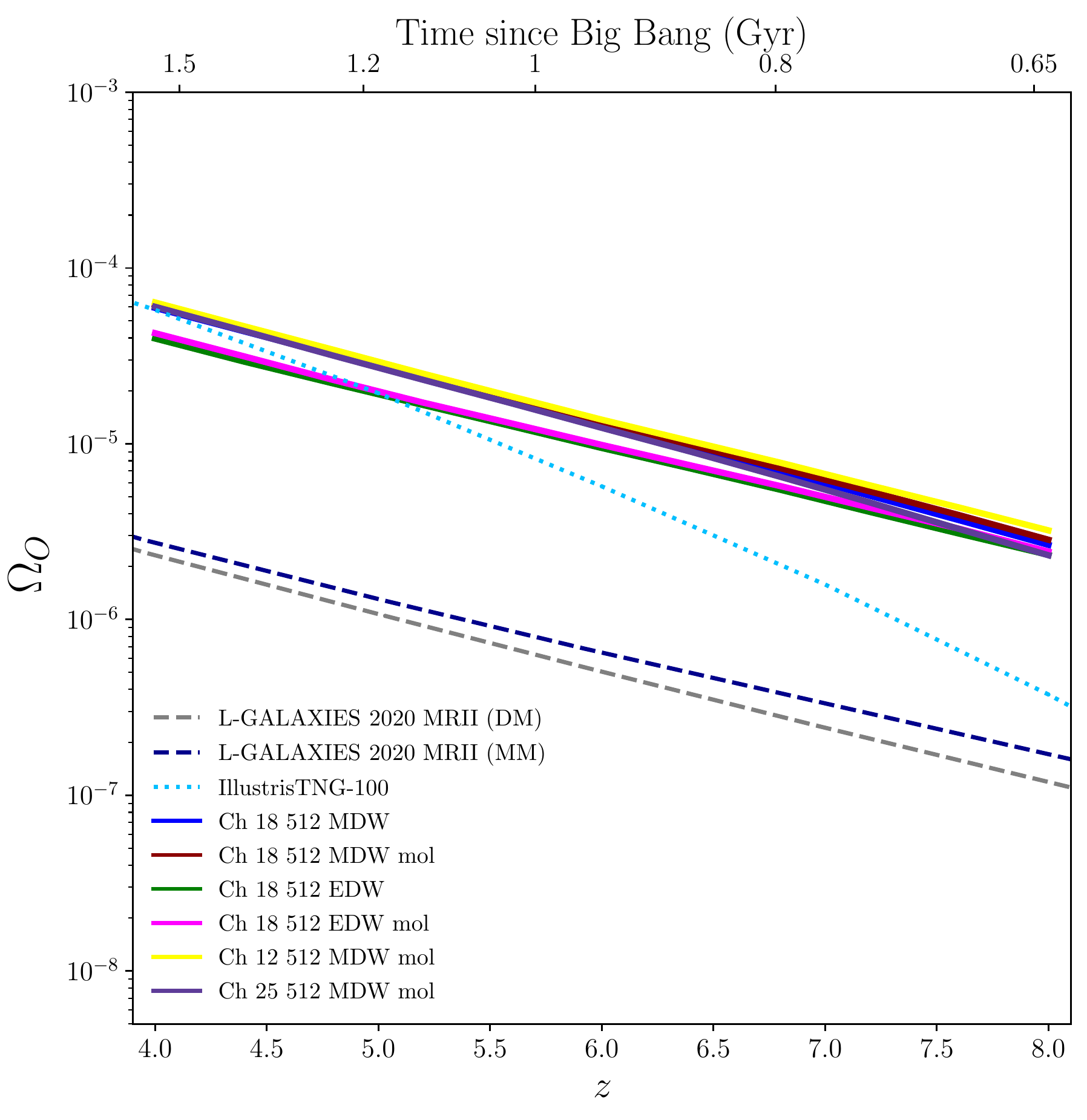}
	\includegraphics[scale=0.45]{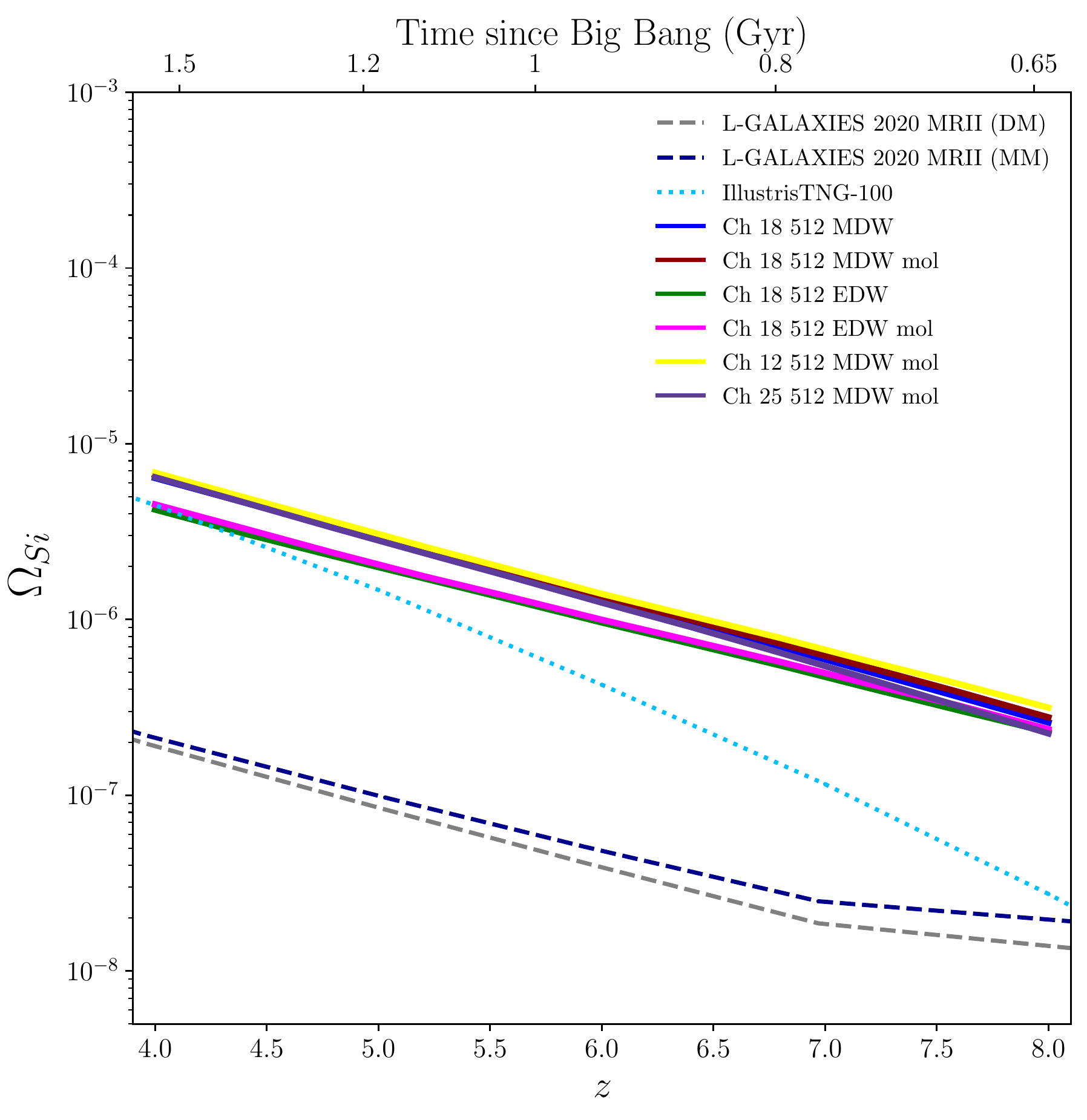}
    \caption{Theoretical predictions for the total cosmological mass densities for Oxygen ($\Omega_{O}$) and Silicon ($\Omega_{Si}$) in the left and right panels, respectively.}
    \label{fig:omegas}
\end{figure}

Detections of absorption lines from \citet{codoreanu2018}, \citet{meyer2019}, and \citet{cooper2019} show that we can set a lower limit for the mass density of Silicon and Oxygen through the reconstruction of the cosmic mass densities of their corresponding metal ions. Besides that, the low-to-high ionization ratio is an independent proxy for the end of Reionization and reveals the gas state in the IGM. Both Oxygen and Silicon have observable ionization states that are exhibited in the spectra of high redshift quasars (OI, SiII and SiIV), redward from the Lyman $\alpha$ emission, and could provide complementary constraints to the metal enrichment, apart from Carbon.\newline
Fig.~\ref{fig:omegas} shows the cosmic evolution of the Oxygen and Silicon mass densities, from $z =$ 8 to $z =$ 4 when the chemical pollution has been occurring for about a billion years in the Universe from stars and supernovae. The cosmic density $\Omega$ as a function of $z$ is obtained by summing the amount of each metal in all gas particles inside the simulated box. Finally, this calculation is divided by the comoving volume. \newline

In addition to the metal mass densities predicted from our models, a comparison with L-Galaxies 2020 for O and Si is presented for their default and modified model \citep[DM and MM, respectively;][]{yates2021b}. Their normalization is similar because the overall amount of cosmic star formation is slightly higher in their modified model than the default setup, despite their distinctive mechanisms introduced to enrich the CGM/IGM; thus, the amount of each element produced overall is similar. For further details on their chemical enrichment modeling, see \citet{yates2013}.
Both models differ in around one order of magnitude from our mass densities due to three main variations among their models and ours: i) their predicted SFRD are lower at all redshifts. Therefore, it is expected that chemical pollution is less effective in this time frame. ii) \citet{yates2021b} assume different metal yields than the ones imposed on our set of simulations. The former is around 0.03 -in order to match late metallicities measured with DLAs-. Instead, the metal yield in all our models is a fixed value of 0.02. iii) Their models include metal outflows released and spread by SNe-II, SNe-Ia, and AGB stars. In our simulations, neither AGB nor AGN are predominant in the feedback mechanisms. Thus fewer processes prevent the outburst of material.\newline

Our predictions are also contrasted against the cosmic mass density calculated with TNG100. Their trends show a much faster evolution than the prediction from L-Galaxies 2020 and ours. These results are consistent with the SFRD exhibited by TNG100, on top of a sophisticated set of stellar yield tables \citep[Table 2 from ][]{pillepich2018}.\newline

From the observational side, a compelling test arises from the evolution of the mean metallicity in the Universe \citep[see Fig.14 in][]{madau2014}, and our Fig.~\ref{fig:omegas}. In \citet{madau2014} an increment of 1 dex to the solar metallicity is presented (under the assumption that the mass of heavy metals per baryon density produced over the cosmic history with a given SFR model and an IMF-averaged yield of $y =$ 0.02), consistent with the predictions from the simulations at $z =$ 4 - 7.\newline

Finally, a correlation between the SFR in the simulations (mainly driven by the feedback processes of the gas) and the cosmic mass densities of these elements is found. Both metals in Fig.~\ref{fig:omegas} show a slight boost at all redshifts when the feedback mechanism is MDW. As mentioned above, the latter feedback prescription is more effective in producing an overcooling of the gas in the CGM, leading to a larger star-formation in the simulations. Hence, more metals are generated and expelled outwards from the galaxies.\newline
It is fair to conclude that the metal pollution scheme that occurs inside the synthetic realizations respects current limits obtained with metal ions of Oxygen and Silicon at high redshift \citet{codoreanu2018}, despite the limited number of absorption lines detected to date.

\section{Discussion and future perspectives}
\label{sec:4}

The numerical models presented in this work show a connection between galaxy properties and their hosting dark matter halos at high redshift, even though the simulations are not state-of-the-art. There is a good agreement of the theoretical models with observational detections of the galaxy stellar mass function at $z =$ 8, 7, and 6, and the cosmic star formation rate at 4 $< z <$ 8. In addition, the numerical runs provide a forecast at high redshift for the halo mass function, the halo occupation fraction, the galaxy stellar-to-halo mass function, and the cosmological mass densities for Oxygen and Silicon. These predictions are concurrent with other theoretical models that account for larger boxes (thus, resolve a broader mass ranges) and/or implement other physical modules, as \textsc{Renaissance}, \textsc{Eagle}, \textsc{IllustrisTNG}, \textsc{UniverseMachine}, L-Galaxies 2020, \textsc{Flares}, \textsc{Astraeus} and \textsc{CROC}. \newline

L-Galaxies 2020 predicts a lower SFR than all our models, and the discrepancy is larger when MDW prescriptions are taken into account because they do not quench the star formation process. This distinction leads to an order of magnitude difference in the cosmic mass densities for Oxygen and Silicon, while comparing both with their default and the modified models of L-Galaxies 2020 and our trends. Different stellar yields and metal enrichment schemes increase the gap between the calculated $\Omega_{X_i}$. On the other hand,  \textsc{TNG100} shows similar outcomes as our predictions, although the latter project has a larger numerical resolution and implements modules with the latest improvements in magneto-hydrodynamical simulations. Now, \textsc{Eagle} simulation was calibrated to reproduce the galaxy stellar mass function and the morphology of galaxies in the local Universe, but it was not meant to be applied at high redshift with just a few resolved galaxies at $z =$ 7. This issue was corrected with \textsc{Flares}, a 3.2 cGpc re-simulated version of \textsc{Eagle}, that accounts for very massive objects during Reionization, that reside in extreme overdensities, not present neither in \textsc{Eagle} nor in our simulations. In that sense, our comparisons of the CSFR with \textsc{Eagle} are more consistent than with \textsc{Flares}.\newline
On the other hand, the models partially differ from the predicted values from \textsc{UniverseMachine}, most likely because the latter set of simulations are run and observational constrained at a vast redshift range (0 $< z <$ 15), covers at least two orders of magnitude more in halo mass and have much more numerical resolution at the galactic level. Instead, the primary motivation for these runs explored was to test the IGM and not explicitly focus on halo properties.\newline

Notably, the reliable cosmic star formation history predicted by the models allows us to have robust theoretical forecasting for chemical pollution. The effective feedback prescriptions play a significant role in regulating and quenching the star formation in the early galaxies and a mechanism to spread out metals to the IGM. \newline

Furthermore, it is worth noting that a big caveat of these theoretical models is that they do not include a module for dust extinction nor low metallicity systems; hence, POPIII are only represented by massive stars, but not for being metal-free in the scheme. As mentioned above, the resolved halo mass range in the simulations is relatively narrow because of the small box sizes.\newline

Besides, these models do not deliver predictions for the faint-end slopes for stellar mass and luminosity functions. Nonetheless, Fig.~\ref{fig:6} is consistent with a constant power-law slope $\alpha =$ -2, with little evolution from $z =$ 8 to 6. The latter result is a key point if one wants to anticipate the future observations from JWST and other large telescopes planned to shed light on the formation of the first structures and the evolution of the Epoch of Reionization.\newline

Moreover, at the redshift range of this study (4 $<z<$ 8), the number of bright galaxies ($M_{\text{UV}} <$ -17 according to the Hubble Space Telescope resolution) account for 40\% of the total amount of the galaxies in the simulations, according to Fig.~\ref{fig:3}, with little evolution in billion years. This effect is due to the significant efficiency of the star formation of high-mass halos. However, the number of massive halos drops in three orders of magnitude in the redshift period from 4 to 8, leading to a decreasing count of bright galaxies, that is consistent with the faint-end slope $\sim$ -2, and findings by \citet{robertson2015}, \citet{liu2016}, \citet{garcia2017a} and \citet{rachana2019}: that Reionization was mainly driven by faint galaxies, due to the small number of bright galaxies in the early Universe. \newline

Finally, it is essential to point out that galaxies and quasars at high redshift generate most ionizing flux that drove the EoR. Although results from \citet{garcia2020} show that variations in the uniform ultraviolet background have little effect on the observed metals, it strongly determines cooling processes and the subsequent star formation/metal pollution. This assumption will be tested once JWST measures the faint end of both the galaxy and quasar luminosity function out to $z \sim$ 10.

\section{Conclusions}
\label{sec:5}

This work presents a set of hydrodynamical simulations at high redshift (4 $ < z < $ 8) with galactic feedback prescriptions and molecular and metal cooling. The study's primary goal is to describe the evolution of galaxy properties and their connection with the dark matter halos that host these galaxies at the tail of Reionization. \newline

The proposed models agree with the observed galaxy stellar mass function at $z =$ 8, 7, and 6, and the cosmic star formation rate at 4 $< z <$ 8. Moreover, they provide a purely theoretical prediction for different galaxy-to-halo statistics and the cosmological mass densities for Oxygen and Silicon during a billion years time-frame. These results are consistent with other simulations that account for modules with diverse physical processes, including \textsc{Renaissance}, \textsc{Astraeus}, CROC and \textsc{UniverseMachine}, that span in more extensive redshift ranges than the ones considered here, and bigger box sizes that allow them to resolve more massive halos, thus, larger galaxies at early times. The best agreement with our models occurs for \textsc{Eagle} and \textsc{TNG100} because these models have similar SPH configurations with modified versions for the galactic winds and equivalent chemical enrichment schemes. L-Galaxies 2020 shows more significant differences in the SFRD and the chemical pollution of CGM and IGM. These contrasting results are mainly driven by introducing a semi-analytical treatment in their case, whereas our models rely on a hydrodynamical set up to describe the physics of the baryons. The more significant discrepancies among our results and other theoretical models appear with the \textsc{UniverseMachine} and \textsc{Flares}. This is due to the large volumes tested by the latter simulations that resolve more massive galaxies. Small boxes as the ones used in this work lead to degraded results in the cosmological scales. However, we remind the reader that our models were initially configured to accurately describe the IGM, at the expense to sacrifice massive structures.\newline

There is a clear correlation between the cosmic star formation history and the metal enrichment of the intergalactic medium in our models, and both processes are regulated by the galaxy and supernova feedback prescriptions in the simulations.\newline
Recovering mass densities of Oxygen and Silicon is a purely theoretical prediction and sets a lower limit that can be contrasted with the observed cosmic mass density from the metal absorption lines visible at high redshift, as OI, SiII, and SiIV.\newline

Finally, the simulations do not provide a direct prediction for the faint-end slope of the galaxy luminosity function, but a constant stellar-to-halo mass ratio and the slope of galaxy stellar mass function in our models lead to an inferred constant power-law slope $\alpha =$ -2, at $z =$ 8 - 6. This last conclusion will be tested and constrained by JWST shortly. The upcoming space and ground-based telescopes will display the assembly of galaxies while the EoR is proceeding and unveil the early Universe with unprecedented precision. 
\section*{Acknowledgments}
The author thanks Universidad ECCI for its support and Swinburne University of Technology where the simulations were run. L.A. Garc\'ia acknowledges the valuable contribution from Robert Yates, Peter Behroozi and Aswin Vijayan for sharing data in private communications and their insightful comments that improved the discussion of the results. In addition, the author recognizes the enormous input to her work from large collaborations such as \textsc{IllustrisTNG} project and \textsc{Flares} by making their simulations products publicly available.

\end{document}